\documentclass[aps,prl,twocolumn]{revtex4-1}
\usepackage{graphicx}
\usepackage{amsfonts}
\usepackage{amsmath}
\usepackage{color}
\usepackage{bm}
\usepackage{amssymb}

\usepackage{multirow}
\usepackage{epstopdf}
\usepackage{natbib}

\newcommand{\beq}{\begin{equation}}
\newcommand{\eeq}{\end{equation}}

\def\colvecnext#1{
        #1
        \global\advance\colveccount-1
        \ifnum\colveccount>0
                \\
                \expandafter\colvecnext
        \else
                \end{pmatrix}
        \fi
}

\begin{document}

\title{
Time-dependent wave packet dynamics calculations of cross sections for ultracold scattering of molecules}
\author{J. Y. Huang,$^{1,2}$  D. H. Zhang}
\affiliation{\scriptsize $^1$State Key Laboratory of Molecular Reaction Dynamics, Dalian Institute of Chemical Physics, Chinese Academy of Science, Dalian 116023, China\\$^2$University of Chinese Academy of Sciences}

\author{R. V. Krems}
\affiliation{Department of Chemistry, University of British Columbia, Vancouver, B.C., V6T 1Z1, Canada}
\date{\today}

\begin{abstract}
Because the de Broglie wavelength of ultracold molecules is very large, the cross sections for collisions of molecules at ultracold temperatures are always computed by the time-independent quantum scattering approach. 
Here, we report the first accurate time-dependent wave packet dynamics calculation for reactive scattering of ultracold molecules. Wave packet dynamics calculations 
can be applied to molecular systems with more dimensions and provide real-time information on the process of bond-rearrangement and/or energy exchange in molecular collisions. 
 Our work thus makes possible the extension of rigorous quantum calculations of ultracold reaction properties to polyatomic molecules 
 and adds a new powerful tool for the study of ultracold chemistry.

\end{abstract}

\maketitle

Cooling molecules to ultracold ($T < 10^{-3}$ Kelvin) temperatures has created a new research field of ultracold molecules, whose applications range from 
tests of fundamental symmetries of nature, to quantum simulation of spin-lattice models, to ultracold chemistry and ultracold dipolar matter \cite{CarrNJP09}. The experiments aimed at the production of ultracold molecules have given rise to new techniques, including the development of high-flux guided molecular beams \cite{MaxwellPRL2005}, chiral-sensitive microwave spectroscopy \cite{PattersonNature2013}, magneto-optical traps for molecules \cite{BarryNature2014}, a molecular fountain \cite{ChengPRL2016}, a molecular synchrotron \cite{HeinerNatPhys2007}, Stark and Zeeman decelerators \cite{BethlemPRL1999,LavertNJP2011}. Central to most experiments in this field are collisions of molecules with atoms or with other molecules. In fact, the most universal method to cool molecules from ambient to ultracold temperatures remains evaporative and/or sympathetic cooling \cite{KetterleAdv1996,StuhlNature2012}. These cooling mechanisms rely on the dominance of momentum transfer in elastic collisions of molecules over inelastic or reactive scattering, which are detrimental to cooling at low temperatures. Theoretical predictions of cross sections for molecular scattering at cold ($\sim 1$ Kelvin) and ultracold temperatures are thus vital for the field of ultracold molecules. 
They are not only necessary for predictions as to which molecular species are amenable to collisional cooling, but also crucial for understanding the broadening mechanisms in precision measurements with trapped molecules, the extent of tunability of microsopic molecular interactions by external fields and the mechanisms of chemical reactions at ultracold tempratures. 

There are generally two rigorous quantum approaches to calculate the cross sections for molecular collisions: the time-independent close coupling (CC) method and the time-dependent wave packet (TDWP) dynamics technique. The CC method represents the eignefunctions of the full time-independent Hamiltonian by a basis set expansion, which reduces the Schr\"{o}dinger equation to a set of coupled differential equations.  All of the previous scattering calculations for collisions of molecules at cold and ultracold temperatures have been done with the CC method or approximate techniques based on the CC method. However, the numerical difficulty of the CC calculations increases as $N^3$ with the number $N$ of the basis states so the CC method is limited to atom - diatom or light molecule - molecule scattering systems. 
The application of the CC calculations to molecule - molecule collisions for heavy molecules or for polyatomic molecules is prohibitively difficult. 
As the field of ultracold molecules is progressing towards polyatomic molecules \cite{ZeppenfeldNature2012,GarrettAngChem2016,KozyryevPRL2017}, it is necessary to extend rigorous quantum calculations of ultracold scattering to larger molecular systems. 
TDWP calculations can be applied to polyatomic molecules. However, until now, TDWP dynamics could not be extended to ultracold temperatures due to the large de Broglie wavelength of ultracold molecules and perceived difficulties with absorbing ultracold molecular wave packets at the boundaries of the calculation grids. 

Here, we overcome these problems and present the first TDWP calculations of cross sections for an ultracold atom - molecule chemical reaction. We illustrate that the TDWP calculations can be extended to the $s$-wave scattering regime and describe properly the threshold behaviour of the reaction cross sections in the limit of vanishing collision energy. We perform calculations for the benchmark F + H$_2$ $\rightarrow$ HF + H reaction, which has been studied widely, both at thermal temperatures and in the ultracold regime \cite{BalakrishnanPRL2001}. We illustrate that the method produces accurate cross sections for reactions of molecules both in the ground state and in excited states, as well as near scattering resonances.

\begin{figure}[!ht]
\includegraphics[width=\columnwidth]{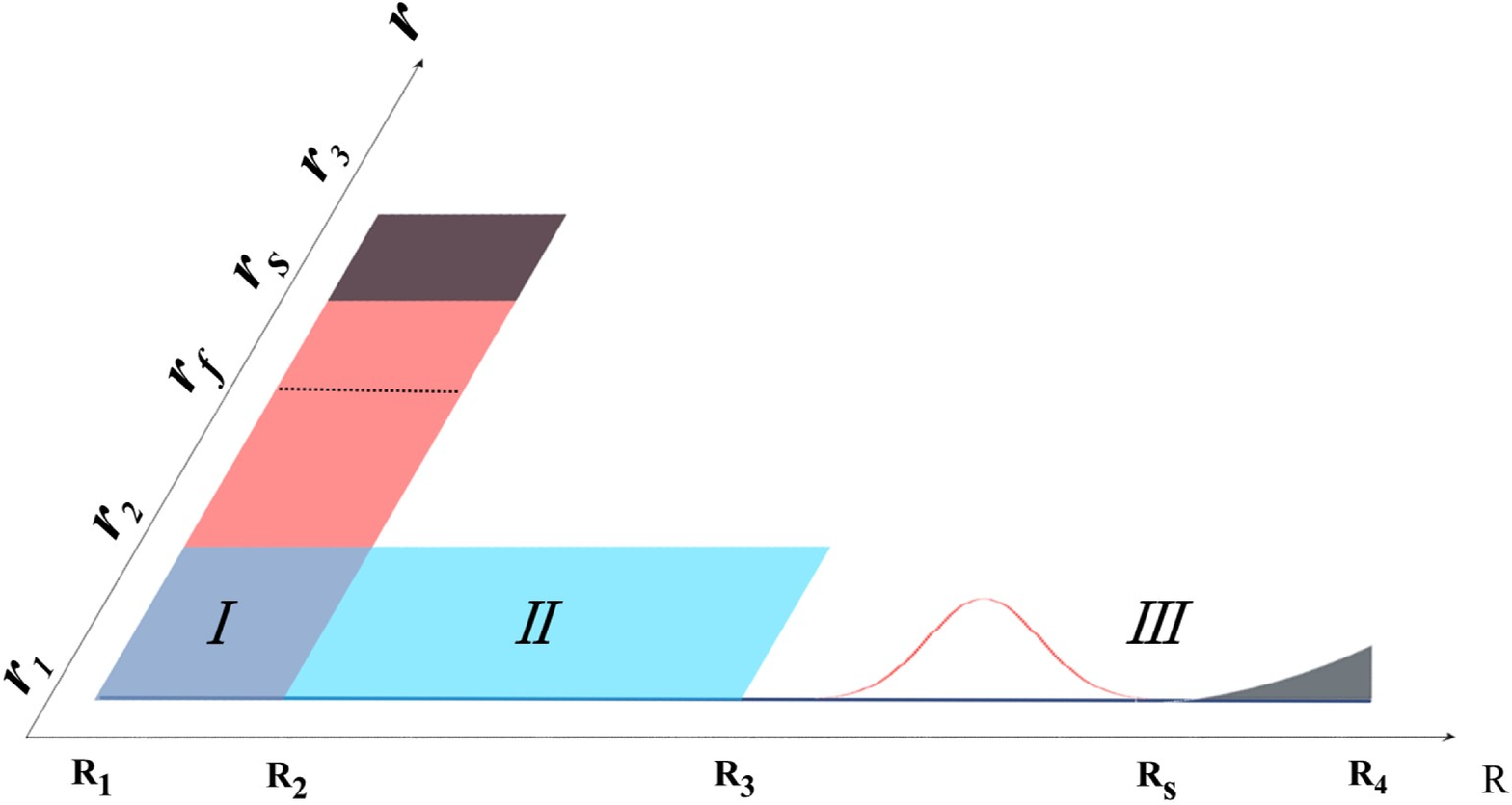}
\caption{\label{fig1} Illustrative drawing of the configuration space for ultracold F + H$_2$ $\rightarrow$ HF + H reaction.  The Roman numeral I denotes the interaction region with the F -- H$_2$ distance $R < R_2$, II denotes the asymptotic region with fewer open channels and III labels the long-range region, where the wave packets are restricted to contain only one molecular state.  The shaded regions show the absorption zones. The reactive flux is evaluated at the surface defined by $r = r_s$.}
\vspace{-7.2cm}
\hspace{2.0cm} \includegraphics[width=0.8in]{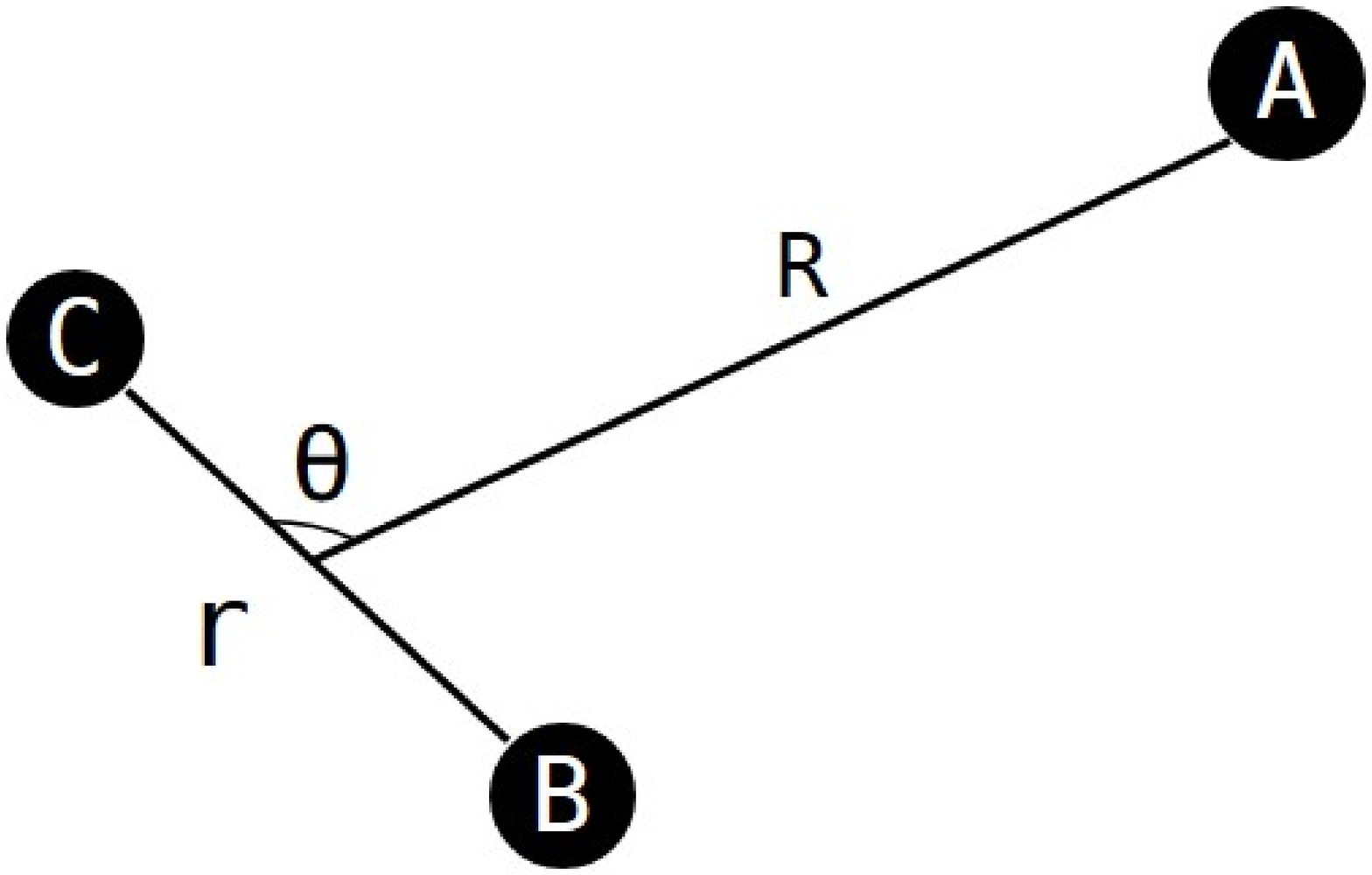}
\vspace{6.2cm}
\end{figure}

{\it Calculation details.}  We solve the time-dependent Schr\"{o}dinger equation with the Hamiltonian 
\begin{equation}
\label{eq:atom3:h3}
\hat{H} = -\frac{\hbar^2}{2 \mu_{R}} \frac{\partial^2}{\partial R^2} + \hat{h}(r) + \frac{(J-j)^2}{2 \mu_{R} R^2}
+\frac{j^2}{2 \mu_{r} r^2} + V(R,r,\theta)
\end{equation} 
in Jacobi coordinates illustrated in Fig.~\ref{fig1}. Here,
$V(R,r,\theta)$ is the atom - molecule interaction potential, 
 $J$ is the total angular momentum of the collision complex, $j$ is the rotational angular momentum of the diatomic molecule
and $\hat h(r)$ is given by
\begin{equation}
\hat{h}(r) = -\frac{\hbar^2}{2 \mu_r} \frac{\partial^2}{\partial r^2} + V_{r}(r),
\end{equation} 
where $V(r)$ is the intramolecular interaction potential. 

 We write the full time-dependent wave function as 
\begin{equation}
\Psi^{JM\varepsilon}(R,r,t) = \sum\limits_{K}D^{J\varepsilon*}_{MK}(\Omega)\psi(t,R,r,\theta;K)
\end{equation}
where $D^{J\varepsilon*}_{MK}(\Omega)$ is the parity-adapted normalized Wigner rotation matrix, depending on the Euler angles $\Omega$, and
$K$ is the projection of $J$ on the body-fixed (BF) quantization axis. The BF states are represented as \cite{suzukiWFexpansion}
\begin{equation}
\psi(t,R,r,\theta,K) = \sum_{n,v, j} F^{K}_{nvj}(t) u^{v}_n(R) \psi_{vj}(r,\theta)
\label{basis-expansion}
\end{equation}
where $\psi_{jv}(r,\theta)$ are the ro-vibrational wave functions of the diatomic molecule in the entrance reaction channel. 
The radial functions $u^{v}_n(R)$ are discussed below.

In the previous work \cite{boss1994H2+OH}, we developed an L-shape wave packet expansion method, which reduces redundant computing of the wave function components for channels with high energy in the asymptotic region, greatly accelerating the TDWP calculations at collision energies $> 0.001$ eV \cite{science2006,science2007,science2010,science2013}. Here, we modify this procedure to apply TDWP calculations to ultracold scattering. 

When an ultracold collision happens, the radial grid explored by the wave packets is extremely extended, which makes general wave packet dynamics calculations prohibitively difficult. 
To make TDWP calculations of ultracold collisions feasible, we develop the following procedure. First, we split the propagation grids into the interaction region (labeled I in Fig.~\ref{fig1}), the asymptotic region (II) and the long-range region ($R > R_3$, labeled III). Second, we split the Hilbert space of molecular states into two subspaces $\cal{Q}$ and $\cal{P}$ spanning variable numbers of states during the propagation. We choose $\cal{Q}$ to include only the initial state in region III, a reduced number of channels (all open channels and a small number of closed channels) in region II and the full set of states needed for converged calculations in region I. The $\cal{P}$ subspace is thus reduced to zero in the interaction region I. 
At any time, we omit the components of the wave packet in $\cal{P}$, which allows us to propagate the wave packet with vanishingly small collision energy to very large distances $R$.  

More specifically, for a molecule initially in the ro-vibrational state ($v_0, j_0$), we restrict the sum over $v$ and $j$ in Eq. (\ref{basis-expansion}) to a single term $\psi_{vj} \Rightarrow \phi_{v_0}^{j_0}(r) Y_{j_0 K}(\theta)$ in region III, a reduced number of terms $\psi_{vj} = \phi_{v}^{j }(r) Y_{j K}(\theta)$ with $v \in [0, v_{\rm as}]$ in region II and all terms $\psi_{vj} = \phi_{v}^{j}(r) Y_{j K}(\theta)$ with $v \in [0, v_{\rm max}]$ in region I. Here, $\phi_v^j(r)$ is the ro-vibrational wave function of the diatomic molecules in the entrance reaction channel and $Y_{j K}(\theta)$ are spherical harmonics. 

The radial functions $u^{v}_n(R)$ are chosen as follows \cite{translationalbasisfunction1993,translationalbasisfunction1994,boss1994H2+OH} :

\begin{equation}
u_n^v = \left\{
\begin{array}{rcl}
\sqrt[]{\frac{2}{R_4-R_1}} \sin \frac{n\pi R}{R_4-R_1} && { v = v_0, j = j_0 }\\
\sqrt[]{\frac{2}{R_3-R_1}} \sin \frac{n\pi R}{R_3-R_1} && { 0 \leq v \leq v_{\rm as}}\\
\sqrt[]{\frac{2}{R_2-R_1}} \sin \frac{n\pi R}{R_2-R_1} && { 0 \leq v \leq v_{\rm max}}\\
\end{array} \right.
\end{equation}

We construct the initial wave packet in the BF representation as 
\begin{equation}
\Psi^{JM\varepsilon}_{v_0j_0K_0} (t = 0) = G(R)\phi_{v_0j_0}(r) \left| JMj_0K_0\varepsilon\right\rangle,
\end{equation}
where $\left| JMj_0K_0\varepsilon\right\rangle$ is the total angular momentum eigenstate in the BF representation with parity of the system $\varepsilon$, $\phi_{v_0j_0}(r)$ is the rovibrational wave function of the diatomic reactant, and $G(R)$ is a Gaussian-shaped function:
\begin{equation}
G(R) = \left (\frac{1}{2\pi\sigma^2} \right )^{1/4}\exp \left [-\frac{{(R-R_0)}^2}{4\sigma^2} - ik_0(R-R_0) \right ]
\end{equation}
describing a wave packet centered at $R_0$, with width $\sigma$ and mean kinetic energy $E_0 = (\hbar/2\mu_R)[k_0^2+\frac{1}{4}\sigma^2]$.

We use the fast sine transform to evaluate the action of the radial Hamiltonian operators on the wave packet. 
The action of the angular kinetic operators on the wave packet is evaluated in a finite basis representation of spherical harmonics. 
The corresponding discrete variable representation \cite{LightDVR} is used to evaluate the action of the potential energy operator in the angular degree of the freedom.
The propagation of the wave functions is computed using the split operator method with a forth-order propagator \cite{4a4Apropagator1995, highordersplitoperator2014}.
 
We need to ensure that the dynamical results are not affected by unphysical reflections from the boundary of the propagation grid.
This is particularly important for an utracold scattering problem involving extremely slow wave packets. This can be achieved by means of an optical potential absorbing the wave packets before they reach the boundary. However, in an ultracold collision, the products of a chemical reaction or inelastic scattering move much faster than the reactants approaching each other with vanishingly low energy.  
Therefore, absorbing potentials must be designed to be different for the initial collision channel and for molecules after the reactive or inelastic scattering. 
In order to prevent reflection from the grid edges, we multiple the wave function by a decaying function $F_{\rm abs}$ near the boundary of the coordinate 
 in each propagation  \cite{Bossabsorbpotential1992,boss1994H2+OH}. 
We set $F_{\rm abs}$ to
\begin{equation}
{F}_{\rm abs} = 
\exp \left [-C_{\rm abs}(x-x_0)/(x_{\rm max}-x_0) \right ]  
\end{equation}
in the interval $x_0 < x < x_{\rm max}$ and $F_{\rm abs} = 1$ otherwise. 
The parameters $x_0$ and $x_{\rm max}$ depend on the collision channel. 
For the products of the chemical reaction, the absorbing potential starts at $x_0 = r_{\text S}$, 
for the products of inelastic scattering -- at $x_0 = R_{\text S}$, and for the initial scattering channel $x_0 = R_3$, with $r_s, R_S$ and $R_3$ 
illustrated in Fig. \ref{fig1}. 




We use a total of $N=2047$ sine basis functions (including 295 for region II and 62 for region I) and the value $R_4= 240$ { a.u.} in the collision energy range $0.1-1$ meV. 
To obtain converged results at lower collision energies, we triple both $N$ and $R_4$ for each order of magnitude of the collision energy decrease. 
We include a total of $v_{\rm max} = 120$ vibrational states for the diatomic molecule fragment in region I, and $v_{\rm as} = 5$ states for region II. For the rotational degree of freedom, we include the spherical harmonics $Y_{jK}$ with $j$ from $0$ up to $j_{\rm max}=90$. The values of the other parameters illustrated in Fig. \ref{fig1} are  $R_1=1$, $r_1=0.6$, $r_3=12$, $r_S = 10$, $R_3 = 35$ a.u. The values of $R_S$ and $R_4$ are chosen to ensure convergence.  


\begin{figure}[!ht]
\includegraphics[width=\columnwidth]{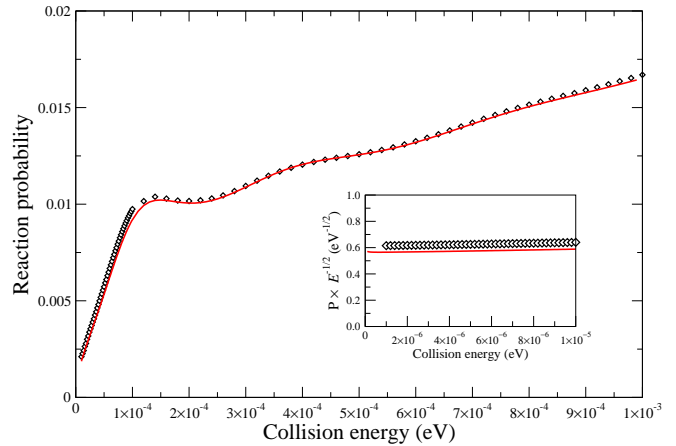}
\caption{\label{fig2} Probability of the chemical reaction F + H$_2(v=0,j=0)$ $\rightarrow$ F + HF summed over all final states of of the reaction products: full line -- time-independent close coupling calculations; symbols - time-dependent wave-packet calculations. The inset shows the low energy reaction probabilities divided by the square root of the collision energy, illustrating the threshold behaviour and the agreement of the two calculations in this limit.}
\end{figure}

{\it Results.} In order to benchmark the performance of the TDWP calculations, we compare the reaction probabilities computed as described above with the results of the time-independent CC calculations. The CC calculations were performed with the ABC code \cite{ABC}, with the same potential energy surface. The integration parameters and the basis sets for the CC calculations were chosen to ensure full convergence. 

Fig. \ref{fig2} shows the comparison of the CC and TDWP results for the reaction of F atoms with H$_2$ molecules in the ground ro-vibrational state in a wide range of energies extending to the ultracold regime. The TDWP calculations reproduce the CC results at all energies, resolving well even the oscillatory behaviour of the reaction probabilities at the collision energy $\sim 2 \times 10^{-4}$ eV.   Even more importantly, the TDWP calculations reproduce the threshold behaviour of the reaction probabilities as the collision energy vanishes. 

As originally shown by Bethe and Placzek \cite{BethePR1937} and Wigner \cite{WignerPR1948}, the probabilities for nuclear reactions vanish as $\propto \sqrt{E}$ when the collision energy $E \rightarrow 0$. It was later shown by Balakrishnan and coworkers \cite{BalakrishnanPRL1998,BalakrishnanCPL2001} that this result also applies to reactive scattering of molecules. Since, at ultracold temperatures, the reaction rate $k$ is related to the reaction probability $P$ as $k \propto P/\sqrt{E} $, the reaction rate is finite and temperature-independent in the limit of zero temperature.  The zero-temperature rate is determined by the value of the reaction probability as it enters the threshold $\propto \sqrt{E}$ regime. Fig. \ref{fig2} shows that the TDWP calculations are accurate all the way down to the threshold regime. There is no need to extend the calculations to lower energies as the reaction probabilities can be extrapolated analytically and the zero temperature rate can be computed based on the value of the reaction probability at $E=10^{-6}$ eV. We thus illustrate that the TDWP calculations describe accurately ultracold reactive scattering. 

In addition to $v_0$ and $j_0$, the initial state of the collision complex is determined by the end-over-end rotational angular momentum $l$. 
Ultracold collisions (of bosons or distinguishable particles) are entirely determined by the components of the wave function with $l = 0$, describing $s$-wave scattering, for which there is no long-range centrifugal barrier to prevent the wave-packet from approaching the reaction region.  It is necessary to verify that TDWP calculations can also accurately describe ultracold scattering with higher partial waves, occurring by tunnelling under the centrifugal barriers.  
To illustrate the accuracy of TDWP calculations for states of higher angular momentum at ultralow energies, we fix the total angular momentum to $J = 0$ and compute the reaction probabilities for H$_2$ in the rotational state $j_0=1$. This is an important case to test, for two reasons. First, this case does not permit $s$-wave scattering so the dominant contribution to the ultracold reaction probability comes from $p$-wave scattering.  
 Second, the reactive scattering of H$_2(j=1)$ with F at ultralow energies is known to be affected by a resonance, which may have a dramatic effect on the threshold behaviour of the reaction probabilities. Since resonances are ubiquitous in ultracold scattering, it is necessary to show that the TDWP calculations are accurate also for resonant scattering.

\begin{figure}[!ht]
\includegraphics[width=\columnwidth]{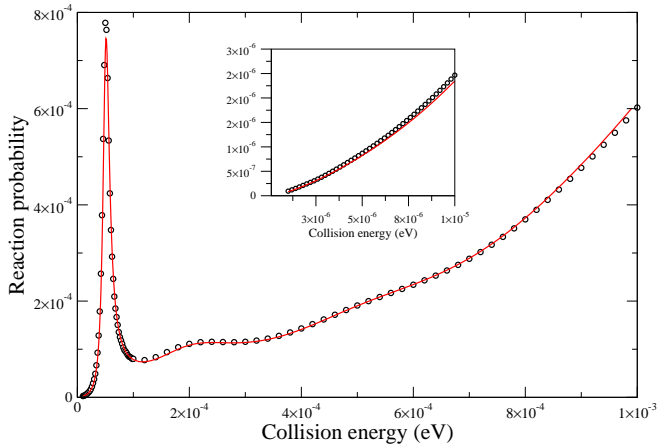}
\caption{\label{fig3} Probability of the chemical reaction F + H$_2(v=0,j=1)$ $\rightarrow$ F + HF summed over all final states of of the reaction products: full line -- time-independent close coupling calculations; symbols - time-dependent wave-packet calculations. The inset shows an enhanced view of the low-energy part of the reaction probability.}
\end{figure}

Fig. \ref{fig3} illustrates the agreement of the TDWP calculations with the CC results for reactions of molecules in the $j=1$ state. The two methods are in excellent agreement for both resonant and threshold reactive scattering. As illustrated by the inset of Fig. \ref{fig3}, the scattering resonance results in a departure of the reaction probabilities from 
the Wigner behaviour at collision energies $>1.5 \times 10^{-5}$ eV. Nevertheless, the TDWP calculations capture the energy dependence of the reaction probabilities accurately, including at the ultralow energies where the threshold energy dependence dominates and at the point of the deviation from the Wigner dependence due to the resonance.

\begin{figure}[!ht]
\includegraphics[width=\columnwidth]{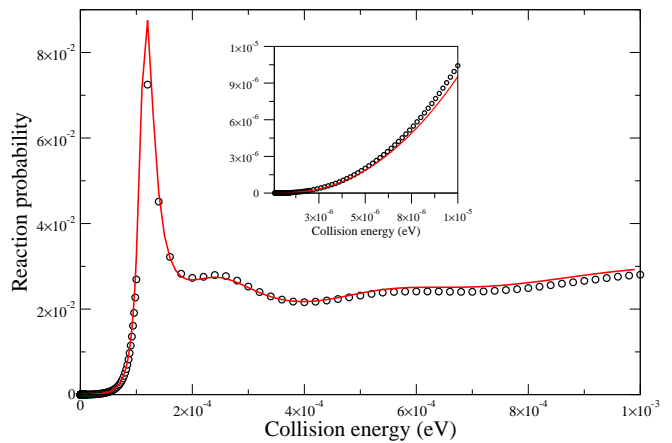}
\caption{\label{fig4} Probability of the chemical reaction F + H$_2(v=1, j=2)$ $\rightarrow$ F + HF summed over all final states of of the reaction products: full line -- time-independent close coupling calculations; symbols - time-dependent wave-packet calculations. The calculations are for $J=0$ so the reaction probabilities shown are determined by $d$-wave scattering in the limit of ultralow collision energy.}
\end{figure}

For molecules in the ground ro-vibrational state, there is only one channel open in regions II and III of Fig. \ref{fig1}. 
Therefore, the results shown in Figs. \ref{fig2} and \ref{fig3} are obtained only with one channel propagated in region III. To verify that the technique described here can be applied also to molecules initially in excited states, we perform TDWP calculations for reaction of H$_2$ in the vibrationally and rotationally excited state $v_0 = 1, j_0 = 2$. 
In this calculation, as described above, we still propagate only one channel in region III, but this channel now corresponds to an excited state, leaving multiple channels energetically accessible at all times. Fig. \ref{fig4} illustrates that this approach produces accurate results in a wide range of energies, including near a resonance. 


{\it Conclusion.} We have illustrated that the time-dependent wave packet dynamics calculations can be extended for the calculations of reaction probabilities of molecules at ultralow collision energies, all the way down to the Wigner threshold regime. Our results show that the reaction probabilities computed with the time-dependent method are accurate both near scattering resonances and in the threshold regime. The time-dependent calculations can be applied to complex (4, 5, and even 6-atoms) systems, which are currently out of reach of time-independent close coupling calculations.  The numerical difficulty of the time-dependent calculations is also similar for abstraction reactions (such as the one considered here) and insertion reactions proceeding through the formation of a strongly bound intermediate reaction complex. By contrast, the time-independent calculations for insertion reactions are much more difficult than the calculations for abstraction reactions. The insertion chemical reactions are particularly important for the research field of ultracold molecules, as most of the ultracold chemistry experiments are performed with alkali metal dimers synthesized from ultracold alkali metal atoms in magneto-optical traps. Alkali metal dimers react predominantly through insertion reactions \cite{BarinovsPRA2008}. Finally, wave packet dynamics calculations offer a powerful method to study ultracold reaction mechanisms by providing real time information on the bond-rearrangement process. 
 Our work thus makes possible the extension of rigorous quantum calculations of ultracold reaction properties to bigger than 2-atom systems and to a variety of experimentally relevant alkali metal dimer systems, and adds a new powerful tool for the study of ultracold chemistry.


\end{document}